\documentclass[12pt,epsf]{article}
\usepackage{graphicx,amsmath,amssymb}
\begin{document}

\def\a{\alpha}
\def\b{\beta}
\def\c{\chi}
\def\d{\delta}
\def\e{\epsilon}
\def\f{\phi}
\def\g{\gamma}
\def\h{\theta}
\def\k{\kappa}
\def\l{\lambda}
\def\m{\mu}
\def\n{\nu}
\def\p{\psi}
\def\q{\partial}
\def\r{\rho}
\def\s{\sigma}
\def\t{\tau}
\def\u{\upsilon}
\def\v{\varphi}
\def\w{\omega}
\def\x{\xi}
\def\y{\eta}
\def\z{\zeta}
\def\D{\Delta}
\def\G{\Gamma}
\def\H{\Theta}
\def\L{\Lambda}
\def\F{\Phi}
\def\P{\Psi}
\def\S{\Sigma}

\def\o{\over}
\def\beq{\begin{eqnarray}}
\def\eeq{\end{eqnarray}}
\newcommand{\gsim}{ \mathop{}_{\textstyle \sim}^{\textstyle >} }
\newcommand{\lsim}{ \mathop{}_{\textstyle \sim}^{\textstyle <} }
\newcommand{\vev}[1]{ \left\langle {#1} \right\rangle }
\newcommand{\bra}[1]{ \langle {#1} | }
\newcommand{\ket}[1]{ | {#1} \rangle }
\newcommand{\EV}{ {\rm eV} }
\newcommand{\KEV}{ {\rm keV} }
\newcommand{\MEV}{ {\rm MeV} }
\newcommand{\GEV}{ {\rm GeV} }
\newcommand{\TEV}{ {\rm TeV} }
\def\diag{\mathop{\rm diag}\nolimits}
\def\Spin{\mathop{\rm Spin}}
\def\SO{\mathop{\rm SO}}
\def\O{\mathop{\rm O}}
\def\SU{\mathop{\rm SU}}
\def\U{\mathop{\rm U}}
\def\Sp{\mathop{\rm Sp}}
\def\SL{\mathop{\rm SL}}
\def\tr{\mathop{\rm tr}}

\newcommand{\bea}{\begin{eqnarray}}   
\newcommand{\eea}{\end{eqnarray}}
\newcommand{\bear}{\begin{array}}  
\newcommand {\eear}{\end{array}}
\newcommand{\bef}{\begin{figure}}  
\newcommand {\eef}{\end{figure}}
\newcommand{\bec}{\begin{center}}  
\newcommand {\eec}{\end{center}}
\newcommand{\non}{\nonumber}  
\newcommand {\eqn}[1]{\beq {#1}\eeq}
\newcommand{\la}{\left\langle}  
\newcommand{\ra}{\right\rangle}
\newcommand{\ds}{\displaystyle}
\def\SEC#1{Sec.~\ref{#1}}
\def\FIG#1{Fig.~\ref{#1}}
\def\EQ#1{Eq.~(\ref{#1})}
\def\EQS#1{Eqs.~(\ref{#1})}
\def\TEV#1{10^{#1}{\rm\,TeV}}
\def\GEV#1{10^{#1}{\rm\,GeV}}
\def\MEV#1{10^{#1}{\rm\,MeV}}
\def\KEV#1{10^{#1}{\rm\,keV}}
\def\lrf#1#2{ \left(\frac{#1}{#2}\right)}
\def\lrfp#1#2#3{ \left(\frac{#1}{#2} \right)^{#3}}
\def\REF#1{Ref.~\cite{#1}}
\newcommand{\Fb}{\bar{\Phi}}
\newcommand{\vbl}{v_{\rm B-L}}
\newcommand{\gbl}{g_{\rm B-L}}
\newcommand{\ubl}{U(1)$_{\rm B-L}$ }
\newcommand{\bl}{{\rm B-L}}


\baselineskip 0.7cm

\begin{titlepage}

\begin{flushright}
TU-902\\
IPMU12-0028\\
UT-12-07
\end{flushright}

\vskip 1.35cm
\begin{center}
{\large \bf 
PeV-scale Supersymmetry from  New Inflation 
}
\vskip 1.2cm
Kazunori Nakayama$^{a,c}$
and
Fuminobu Takahashi$^{b,c}$

\vskip 0.4cm

{\it $^a$Department of Physics, University of Tokyo, Tokyo 113-0033, Japan}\\
{\it $^b$Department of Physics, Tohoku University, Sendai 980-8578, Japan}\\
{\it $^c$Institute for the Physics and Mathematics of the Universe,
University of Tokyo, Kashiwa 277-8568, Japan}\\

\vskip 1.5cm

\abstract{ We show that heavy supersymmetric particles around
  $O(100)$\,TeV to $O(1)$\,PeV naturally appear in new inflation in
  which the Higgs boson responsible for the breaking of U(1)$_{\rm
    B-L}$ plays the role of inflaton.  Most important, the
  supersymmetric breaking scale is bounded above by the inflationary
  dynamics, in order to suppress the Coleman-Weinberg potential which
  would otherwise spoil the slow-roll inflation.  Our scenario has
  rich phenomenological and cosmological implications: the Higgs boson
  mass at around $125$\,GeV can be easily explained, non-thermal
  leptogenesis works automatically, the gravitino production from
  inflaton decay is suppressed, the dark matter is either the lightest
  neutralino or the QCD axion, and the upper bound on the inflation
  scale for the modulus stabilization can be marginally satisfied.  }
\end{center}
\end{titlepage}

\setcounter{page}{2}

\section{Introduction}
\label{sec:1}
%
The concept of symmetry has been a guiding principle in modern
physics. For instance, the structure of the standard model (SM) is
dictated by the SM gauge symmetries, $SU(3)_C \times SU(2)_L \times
U(1)_Y$. The central issue is then how to break symmetry, because
clearly we are living in a broken phase: the observed rich structure
in our Universe would be impossible in a completely symmetric
vacuum. In the celebrated Higgs mechanism~\cite{Higgs:1964pj}, gauge
symmetry is spontaneously broken by a vacuum expectation value (VEV)
of a Higgs field.

Recently the ATLAS and CMS collaborations have provided hints for the
existence of a SM-like Higgs particle with mass of about
$125$\,GeV~\cite{Higgs-LHC}.  The relatively light Higgs boson mass
suggests the presence of new physics at scales below the Planck
scale~\cite{EliasMiro:2011aa}.  In a supersymmetric (SUSY) extension
of the SM (SSM), the Higgs boson mass can be explained if the typical
sparticle mass is at $O(10)$\,TeV or heavier.  This casts doubt on the
conventional naturalness argument as the correct guiding principle for
understanding the physics at and beyond the weak scale.

Once the existence of the SM-like Higgs boson is confirmed, it would
immediately mean that the Higgs mechanism is indeed realized in
nature, and some other gauge symmetries may be broken in a similar
manner. Those symmetries may have been restored in the past because
the Universe was much hotter and denser at early times.  Thus probably
our Universe experienced a series of phase transitions in course of
its evolution.

The inflationary paradigm has been well established so
far~\cite{Guth:1980zm}. Despite its great success, it is not yet known
what the inflaton is.  It is natural to expect that one of the Higgs
fields which trigger phase transitions in the early Universe is
responsible for the inflation. Indeed, this possibility was
extensively discussed in the early 80's under the name of new
inflation~\cite{Linde:1981mu}.  The phase transition in the new
inflation was of Coleman-Weinberg (CW) type~\cite{Coleman:1973jx},
where the inflaton was the Grand Unification Theory (GUT) Higgs boson with the mass at the origin
being set to be zero. Although this scenario was very attractive, it
was soon realized that the CW correction arising from the gauge boson
loop makes the inflaton potential too steep to produce the density
perturbation of the correct magnitude, $\delta \rho/\rho \sim
10^{-5}$~\cite{Starobinsky:1982ee}.  One solution was to consider a
gauge singlet inflaton, which has extremely weak interactions with the
SM particles.  Although the inflation model may lose its connection to
the GUT in this case, such gauge singlets are ubiquitous in the string
theory, and so, one of them may be responsible for the inflation.
Another way to resolve the problem was to introduce
SUSY~\cite{Ellis:1982ed}. Then the CW potential becomes suppressed
because of the cancellation among bosonic and fermionic degrees of
freedom running the loop.

Recently the present authors proposed a new inflation model in which a
Higgs field responsible for the breaking of \ubl symmetry plays the
role of inflaton~\cite{Nakayama:2011ri}.  It was found that the SUSY
must be a good symmetry at scales below the Hubble parameter during
inflation.\footnote{ In Ref.~\cite{Ellis:1982ed}, the soft SUSY
  breaking mass was (implicitly) assumed to scale in proportion to the
  GUT Higgs boson.  In their Eq.~(8), the dependence of the CW
  potential on the GUT Higgs boson was factored out, and then they
  substituted the mass splitting relation Eq. (21) into Eq.~(8). In
  effect, this is equivalent to assuming that the soft SUSY breaking
  mass is proportional to the GUT Higgs boson VEV.  Therefore the
  upper bound on the SUSY breaking was overestimated, and it was
  actually higher than the Hubble parameter during inflation, which
  clearly does not make sense, because it would mean that there is no
  SUSY at the inflation scale.  To our knowledge, this error was not
  corrected until Ref.\cite{Nakayama:2011ri}. } Interestingly, we
obtained an upper bound on the soft SUSY breaking mass, $\tilde{m}
\lesssim O(10){\rm\,TeV} - O(1)$\,PeV for the \ubl breaking scale of
$\GEV{15}$ inferred from the neutrino oscillation
data~\cite{Nakayama:2011ri}. Furthermore, the inflaton predominantly
decays into a pair of right-handed neutrinos, and non-thermal
leptogeness~\cite{Lazarides:1991wu,Asaka:1999yd} works almost
automatically. The implication for the SM-like Higgs boson mass in
this framework was studied in Ref.~\cite{Nakayama:2011ys}.

In this paper, we study the inflationary dynamics of the \ubl new
inflation as well as its subsequent thermal history of the \ubl new
inflation in detail.  The spectral index is calculated with a greater
accuracy and found to be perfectly consistent with the current WMAP
data, $n_s \simeq 0.968 \pm 0.012$~\cite{Komatsu:2010fb}. In
particular we will see that the CW potential will play an important
role to increase $n_s$ to provide a better fit to the WMAP data.  We
will consider the implication of the SUSY breaking mass from the
inflationary dynamics for the SM-like Higgs boson mass. We also
discuss various phenomenological and cosmological implications such as
non-thermal leptogenesis, gravitino production from inflaton decay,
dark matter (DM), the Polonyi problem, and the modulus destabilization
problem~\cite{Kallosh:2004yh}.  It is noteworthy that in such a
minimal extension of the SSM, the observed data such as the spectral
index of the density perturbation and the SM-like Higgs boson mass can
be explained while naturally creating the right amount of the baryon
asymmetry without the gravitino and Polonyi problems.

 \vspace{5mm}

 The rest of the paper is organized as follows. We will briefly review
 how the SUSY breaking is bounded from above in the new inflation in
 Sec.~\ref{sec:2}, and derive an important upper bound on the soft
 SUSY breaking mass. In Sec.~\ref{sec:3} we discuss the dynamics of
 the U(1)$_{\rm B-L}$ new inflation in detail. In Sec.~\ref{sec:4} we
 discuss the reheating of the inflaton. The implications for the
 SM-like Higgs boson mass is discussed in Sec.~\ref{sec:5}.  We
 discuss various implications of our scenario in Sec.~\ref{sec:6}. The
 last section is devoted for conclusions.

\section{Upper bound on SUSY breaking}
\label{sec:2}
Let us briefly review how the SUSY breaking is bounded above for the
successful new inflation using a gauge non-singlet inflaton, following
Ref.~\cite{Nakayama:2011ri}.  The bound essentially comes from the
requirement that the radiative correction to the inflation potential
should be suppressed since otherwise the slow-roll inflation would not
last long enough and the density perturbation would be too large.

Consider a Higgs boson $\varphi$ responsible for the breaking of
U(1)$_{\rm B-L}$ symmetry.  In the new inflationary scenario, the
inflaton sits near the origin at the beginning of inflation. If the
inflaton potential is sufficiently flat around the origin, the
inflation takes place. As $\varphi$ is charged under the \ubl
symmetry, the inflaton potential receives a radiative correction from
the gauge boson loop.  The general form of the CW effective potential
is given by~\cite{Coleman:1973jx}
\begin{equation}
V_{\rm CW}\; =\; \frac{1}{64\pi^2}
\sum_i (2 S + 1) (-1)^{2S} M_i^4(\varphi) \ln\lrf{M_i^2(\varphi)}{\mu^2},
  \label{CW}
\end{equation}
where $\mu$ is the renormalization scale, and 
 the mass eigenvalues of the particles coupled to $\varphi$ are represented by $M_i (\varphi)$.
Since the mass of the U(1)$_{\rm B-L}$ gauge boson is given by $m_{\rm
  GB} = \sqrt{2} \gbl q_\varphi \la \varphi \ra$, the inflaton
potential receives the CW correction as
\beq
V_{\rm CW, gauge}(\s) \;=\; 
\frac{3}{64\pi^2} \gbl^4 q_\varphi^4  \s^4\, \ln\left(\frac{\gbl^2 q_\v^2  \s^2}{\mu^2} \right),
  \label{CW1}
\eeq
where $\gbl$ represents the gauge coupling of U(1)$_{\rm B-L}$,
$q_\varphi$ is the \ubl charge of $\varphi$, and $\sigma$ denotes the
radial component of $\varphi$, $\sigma \equiv \sqrt{2} |\varphi|$.

It is well known that the CW potential arising from the gauge boson
loop makes the effective potential so steep that the resultant density
perturbation becomes much larger than the observed
one~\cite{Starobinsky:1982ee}.  One plausible way to solve the problem
is to introduce SUSY~\cite{Ellis:1982ed}.  In the exact SUSY limit,
contributions from boson loops and fermion loops are exactly canceled
out.  However, if SUSY is broken, we are left with non-vanishing CW
corrections, which are estimated below.

In SUSY, two U(1)$_{\rm B-L}$ Higgs bosons are required for anomaly
cancellation.  Let us denote the corresponding superfields as
$\Phi(+2)$ and ${\bar \Phi}(-2)$ where the number in the parenthesis
denotes their B$-$L charge.  The $D$-term potential vanishes along the
$D$-flat direction $\Phi {\bar \Phi}$, which is to be identified with
the inflaton. Actually, a linear combination of the lowest components
of $\Phi$ and $\bar{\Phi}$ corresponds to $\v$. We can simply relate
$\Phi$ and $\Fb$ to $\v$ as $|\Phi| = |\Fb| = |\v|/\sqrt{2}$. The \ubl
charge of $\v$ is set to be $q_\v = 2$ in the following.

The gauge boson has mass of $m_{S}^2 = \gbl^2 q_\v^2 \s^2$, where $\sigma
\equiv \sqrt{2} |\v|$.  On the other hand, there are additional
fermionic degrees of freedom, the U(1)$_{\rm B-L}$ gaugino and
higgsino, whose mass eigenvalues are given by $m_F = \gbl q_\v \s \pm
\frac{1}{2} M_\lambda$, where $M_\lambda$ denotes the soft SUSY
breaking mass for the \ubl gaugino.  Because of the SUSY breaking mass
$M_\lambda$, the CW potential does not vanish and the inflaton
receives a non-zero correction to its potential.  Inserting the field
dependent masses into the CW potential (\ref{CW}), and expanding it by
$M_\lambda/ (g q_\v \s)$, we find
\begin{equation}
	V_{\rm CW, gauge}^{\rm susy}(\s) \simeq - \frac{3 \gbl^2}{8\pi^2}
	\lrfp{q_\v}{2}{2} M_\lambda^2 \s^2
	\ln \left(\frac{\gbl^2 q_\v^2 \s^2}{\mu^2} \right),
	\label{CW2}
\end{equation}
where we have also taken into account of the inflaton as well as the scalar perpendicular to
the D-flat direction. 
Thus, in the presence of SUSY, the CW potential becomes partially
canceled and the dependence of the inflaton field has changed from
quartic to quadratic as long as $M_\lambda \ll \gbl q_\v \sigma$, in
contrast to the result of Ref.~\cite{Ellis:1982ed}.  Note that the
correction still contains a logarithmic factor, which may not be
negligible if we consider the whole evolution of the inflaton.

For successful inflation, we require the curvature of the CW potential
(\ref{CW2}) to be at least one order of magnitude smaller than $H_{\rm
  inf}$ for $\s \lesssim \s_{\rm end}$.  Here
  $H_{\rm  inf}$ is the Hubble parameter during inflation, and 
   $\s_{\rm end}$ is the point where the slow-roll
condition breaks down and the inflation ends.  Therefore, we obtain
the following constraint on the soft SUSY breaking mass for the \ubl
gaugino:
\begin{equation}
	\gbl M_\lambda \; \lsim \;  O(0.1) H_{\rm inf}.
	\label{bound}
\end{equation}
For the gauge coupling of order unity, this bound reads 
$M_\lambda \; \lsim \;  O(0.1) H_{\rm inf}$.

The \ubl Higgs boson is also coupled to the right-handed neutrinos to
give a large Majorana mass. We consider the following interaction,
\beq
-{\cal L} =\sum_i \frac{y_{\v, i}}{2}  \varphi \,{\bar {\nu}_{R,i}^c} \nu_{R,i} + {\rm h.c.},
\label{MN-nonSUSY}
\eeq
where the subscript $i$ represents the generation. The right-handed
neutrino mass is given by $M_{N,i} = y_{\v,i} \sigma/ \sqrt{2}$. This
interaction similarly contributes to the CW potential as\footnote{ In
  principle it is possible to cancel $V_{\rm CW, gauge}$ with $V_{\rm
    CW, N}$ by fine-tuning the Yukawa coupling $y_{\v i}$. In this
  case, the successful inflation takes place without SUSY, and the
  above upper bound on the soft SUSY breaking mass does not hold. We
  do not consider this case further in this paper. }
\beq
V_{\rm CW, N} \;=\; - \sum_i\frac{1}{8 \pi^2} y_{\v,i}^4 \s^4 \,\ln{\left(\frac{ 2 y_{\v,i}^2 \sigma^2}{\mu^2} \right)}.
	\label{CW3}
\eeq
This can similarly spoil the inflationary dynamics. In the presence of
SUSY, there are right-handed sneutrinos.  Let us write its mass as
$M_{\tilde{N},i}^2 = ( \sqrt{2} y_{\v,i} \sigma)^2 + m_{\tilde
  N,i}^2$, where $m_{\tilde N,i}^2$ represents the soft SUSY breaking
mass for the right-handed sneutrinos.  The CW potential is then
\beq
V_{\rm CW, N}^{\rm susy}(\s) \;=\; \sum_i \frac{y_{\v,i}^2}{8 \pi^2}  m_{\tilde{N},i}^2 \s^2 
\ln \left(\frac{2 y_{\v,i}^2  \s^2}{\mu^2} \right).
 	\label{CW4}
\eeq
For  successful inflation, the soft mass is bounded above as before:
\beq
\sqrt{\sum_i y_{\v,i}^2 m_{\tilde N,i}^2} \;\lesssim\;  O(0.1) H_{\rm inf}.
	\label{bound2}
\eeq
If the Yukawa coupling for the heaviest right-handed neutrino
$\nu_{R,3}$ is of order unity, the bound reads $m_{\tilde N,3}
\;\lesssim\; O(0.1) H_{\rm inf}$.

In the gravity mediation, $M_\lambda$, $m_{\tilde N,i}$ as well as the
soft SUSY masses for the SSM particles are considered to be comparable
to the gravitino mass $m_{3/2}$.  On the other hand, in anomaly
mediation~\cite{Giudice:1998xp}, they may be suppressed compared to
the gravitino mass, but for a generic form of the K\"ahler potential,
$m_{\tilde N}$ and the sfermion masses are comparable to the gravitino
mass.  We assume the latter case when we consider the case of anomaly
mediation.  On the other hand, in the gauge mediation, the relation
between the soft masses and the gravitino mass is model-dependent, and
we do not consider gauge mediation in this paper.

The inflation places a robust upper bound on the soft SUSY breaking parameter
of the \ubl gaugino and the right-handed sneutrino.  In particular, for
$\gbl$ and $\sqrt{\sum_i y_{\v,i}^2}$ of order unity, both $M_\lambda$
and $m_{\tilde N,i}$ should be smaller than $H_{\rm
  inf}$. Furthermore, as long as $M_\lambda$ and $m_{\tilde N,i}$ are
comparable to the soft SUSY breaking mass for the SSM particles
${\tilde m}$ as in the gravity or anomaly mediation we obtain
\beq
       	      {\tilde m} \;\lesssim\; O(0.1) H_{\rm inf},        
	      \label{mH}  
\eeq
which relates the inflation scale to the SUSY breaking.\footnote{
\label{ftn}
$\tilde{m}$ should be considered as representing the sfermion mass, if
the gaugino mass is suppressed as in the anomaly mediation.  }  As we
shall see later, the inflation scale varies from $10^6$\,GeV ($n=2$)
to $\GEV{10}$ or heavier ($n\geq3$).  (See \EQ{KW} for the definition
of the power $n$.)  We will focus on the simplest case of $n=2$,
because it provides an interesting upper bound on ${\tilde m}$ and
because the VEV of the inflaton is very close to the see-saw scale
$\sim \GEV{15}$ suggested by the neutrino oscillation data.  We shall
see that in the case of $n\geq 3$ some of the nice features of the
model are preserved, although the direct connection between the
inflaton VEV and the see-saw scale is lost.

We emphasize here that this novel bound on the soft SUSY breaking mass
is derived from the requirement that the inflation should occur.  Even
if high-scale SUSY breaking scale is favored in the string landscape,
the anthropic pressure by the inflation may constrain the SUSY
breaking scale to be below the inflation scale.  Also, in this case we
have a prediction that the SUSY breaking scale should be close to the
inflation scale.  We assume that this is the case, because, if it is
biased to lower SUSY breaking scale, we should have already seen SUSY
particles at the collider experiments. Interestingly, as we will see,
the observed value of the scalar spectral index even suggests that the
upper bound is saturated.  Even if the SUSY particles are too heavy to
be discovered at the LHC, we may be able to see the hint for the SUSY
breaking scale much higher than the electroweak breaking from the
large radiative correction to the SM Higgs boson
mass~\cite{Giudice:2011cg}. We will come back to this issue in
Sec.~\ref{sec:5}.

Lastly let us mention the applicability of the inequality (\ref{mH}).  As is clear from
the derivation,  the upper bound on SUSY breaking derived applies to any inflation
models in which there are fields coupled to the inflaton with a coupling of order unity,
and they have inflaton-dependent mass. In particular, this is the case if the inflaton is 
charged under gauge symmetry or if the inflaton has a Yukawa coupling with fermions,
as we have seen above. Note that it is applicable to the gauge-singlet inflation 
models, if the inflaton has a sizable Yukawa coupling like (\ref{MN-nonSUSY}).

\section{U(1)$_{\rm B-L}$ New Inflation}
\label{sec:3}
In the previous section we have seen that the SUSY must be a good
symmetry at the inflation scale. Therefore the inflation sector can be
described in a supersymmetric Lagrangian.

The K\"ahler and super-potentials for the inflation are given by~\cite{Asaka:1999jb}
\bea
\label{KW0}
K &=& |\F|^2 + |\Fb|^2 + |\c|^2 + \frac{k_1 }{4} |\F|^4 +  \frac{k_2}{4}  |\Fb|^4
\\ \non && 
+ k_3 |\F|^2 |\c|^2 + k_4 |\Fb|^2 |\c|^2 + \frac{1}{4} k_5 |\c|^4+ \cdots,\\
W &=& \chi \left(v^2 - g (\F \Fb)^n \right),
\label{KW}
\eea
where $k_i$ ($i = 1 - 5$) and $g$ represent a coupling constant of
order unity, $\cdots$ denotes higher order terms and we adopt the
Planck unit, $M_p \approx 2.4 \times \GEV{18} = 1$.  The charge
assignment of $\F$, $\Fb$ and $\c$ are shown in Table~\ref{tab}.  Note
that we have introduced a discrete $Z_n$ symmetry under which only
$\Fb$ is charged. Such a discrete symmetry is necessary to ensure a
flat potential for the inflaton.

The \ubl and other symmetries may be restored in the early Universe,
because of the thermal mass and/or the Hubble-induced mass. If so, the
origin $\F = \Fb = 0$ is chosen as the initial condition. As the
Universe expands, the temperature and the Hubble parameter decrease,
and finally the inflation takes place when the inflaton potential
dominates the energy density of the Universe, if the inflaton
potential is sufficiently flat.

The CW potential, which could spoil the slow-roll inflation, can be
sufficiently suppressed if the typical SUSY breaking mass of the \ubl
gaugino and the right-handed neutrino is (much) smaller than the
inflation scale.  We assume that this is the case for the moment and
consider the supersymmetric part of the inflaton potential. We shall
discuss the effect of the CW potential ((\ref{CW2}) and (\ref{CW4}))
on the inflation dynamics, especially on the spectral index $n_s$,
later in this section.  The effect of a constant term in the
superpotential was studied in Ref.~\cite{Nakayama:2011ri}; assuming
$|k_5| = O(1)$, it was found that the inflaton dynamics is not
affected as long as $m_{3/2} \lesssim O(0.1) H_{\rm inf}$, which is
similar to \EQ{mH}. We will come back to this issue in
Sec.~\ref{sec:6}.

\begin{table}[t]
\begin{center}
\begin{tabular}{c|c|c|c}
 &$\F$ & $\Fb$ & $\c$ \\ \hline \hline 
\ubl & -2 & 2& 0 \\
U(1)$_R$ & 0&0&2 \\
$Z_n$ & 0 & 1 & 0
\end{tabular}
\end{center}
\caption{The charge assignment for $\F$, $\Fb$ and $\c$.}
\label{tab}
\end{table}%

For a field value greater than the Hubble parameter during inflation,
the D-term potential forces $\F$ and $\Fb$ to be along the D-flat direction,
$|\F| = |\Fb| = \frac{1}{\sqrt{2}} |\v|$, where $\v$ is a complex scalar field.
Focusing on the radial component, $\sigma \equiv \sqrt{2} |\v|$,
the scalar potential is approximately given by
\beq
V(\sigma, \chi) \;\simeq\; v^4  - \frac{1}{2} \left(\frac{k_3 + k_4-2}{4} \right) v^4  \sigma^2
- \frac{g }{2^{2n-1}} v^2 \sigma^{2n} + \frac{g^2}{2^{4n}} \sigma^{4 n} 
- k_5 v^4 |\c|^2.
\eeq
We assume $k_5 < -\frac{3}{4}$ so that $\chi$ is stabilized at the
origin during inflation. In order for the slow-roll inflation to take
place, we also require the inflaton mass term is much smaller than the
Hubble parameter,
\beq
k \;\equiv\; \frac{k_3 + k_4-2}{4} \lesssim O(0.01).
\eeq
The tuning of the inflaton mass is known as the $\eta$-problem. We do
not care about this fine-tuning at the level of $1\%$, because it can
be easily compensated by the subsequent exponential expansion and
because perhaps we cannot live in an Universe which has not
experienced inflation. We note that, in general, $k_3$ and $k_4$ do
not have to be small, and we expect them to be of order
unity.\footnote{ Since either $k_3$ or $k_4$ is likely greater than
  unity, either $\Phi$ or $\Fb$ acquires a tachyonic mass in the
  vicinity of the origin, developing a local minimum.  This may make
  the eternal inflation more likely.  } The inflation dynamics in this
model is same as in the single-field new inflation
model~\cite{Izawa:1996dv}, which was studied in detail in
Ref.~\cite{Ibe:2006fs}.

Let us rewrite the inflaton potential, assuming $\chi$ is stabilized
at the origin:
\beq
V(\sigma) \;\simeq\; v^4  - \frac{1}{2} k v^4  \sigma^2
- \frac{g }{2^{2n-1}} v^2 \sigma^{2n} + \frac{g^2}{2^{4n}} \sigma^{4 n}.  
\label{Vinf}
\eeq
After inflation, the inflaton $\sigma$ is stabilized at the potential
minimum given by
\beq
\sigma_{\rm min} \;\simeq\; 2 \lrfp{v^2}{g}{\frac{1}{2n}}.
\eeq
The U(1)$_{\rm B-L}$ symmetry is spontaneously broken by the inflaton
vev.  We define the breaking scale as
\beq
v_{\rm B-L} \;\equiv\; \frac{\sigma_{\rm min}}{\sqrt{2}} \ = \sqrt{2}\lrfp{v^2}{g}{\frac{1}{2n}}.
\eeq
Note that $\vbl$ cannot take an arbitrary value because the coupling
$g$ should not be much larger than $O(1)$ for the K\"ahler potential
\EQ{KW0} to be valid.

In order to estimate the Hubble parameter during inflation, we need to
solve the inflation dynamics and estimate the density perturbation.
When the inflaton sits near the origin, the slow-roll inflation takes
place.  As the inflaton rolls down on the potential, the curvature of
the potential becomes gradually non-negligible, and finally the
slow-roll inflation ends when one of the slow-roll parameters, $\eta$,
becomes order unity.  The $\eta$ is given by
\beq
\eta \;\equiv\; \frac{V''(\s)}{V(\s)} \simeq -k - \frac{n (2n-1)g}{2^{2(n-1)} v^2} \s^{2(n-1)},
\eeq
and $|\eta|$ becomes unity at $\s = \s_{\rm end}$, which is given by
\beq
\s_{\rm end} \;\approx\; 2 \lrfp{(1-k) v^2}{n(2n-1) g}{\frac{1}{2(n-1)}}.
\eeq

\begin{figure}[t]
\begin{center}
\includegraphics[scale=0.6]{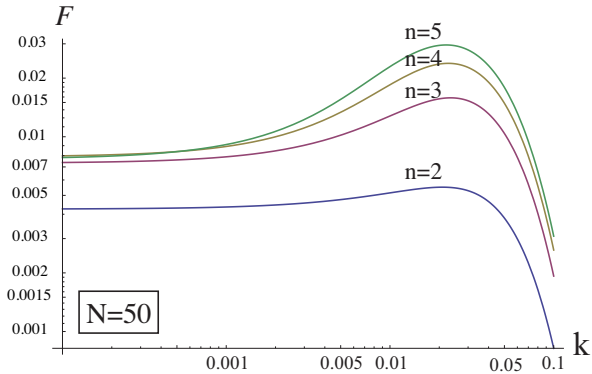}
\caption{The behavior of the function ${\cal F}(k,n, N)$, where we set $N=50$.}
\label{f}
\end{center}
\end{figure}

Under the slow-roll approximation, the equation of motion for the inflaton
is given by
\beq
3 H \frac{d \s}{dt} + V'(\s) \approx 0,
\eeq
or equivalently
\beq
3 H^2 \frac{d \s}{dN} + V'(\s) \approx 0,
\eeq
where $N$ denotes the e-folding number. Solving this equation of motion
we obtain
\beq
\sigma(N) \;\approx\; 2 \lrfp{kv^2}{ng}{\frac{1}{2(n-1)}} {\cal G}(k,n,N)^{-\frac{1}{2(n-1)}},
\eeq
where we have defined
\beq
{\cal G}(k,n,N) \;\equiv\; e^{2k(n-1)N} \left(1+\frac{k}{1-k} (2n-1)\right)-1.
\eeq
The curvature perturbation can be expressed in terms of the inflaton potential,
\beq
\Delta_{\cal R}^2 \;=\; \frac{1}{12 \pi^2} \frac{V(\s)^3}{V'(\s)^2} = (2.430 \pm 0.091)\times 10^{-9},
\label{WMAP_norm}
\eeq
where we have used the WMAP normalization in the second
equality~\cite{Komatsu:2010fb}.  The Hubble parameter during inflation
is given by
\beq H_{\rm inf} \;\simeq\; \frac{v^2}{\sqrt{3}} \simeq
\sqrt{\Delta_{\cal R}^2} \, {\cal F}(k, n, N)\,v_{\rm
  B-L}^\frac{n}{n-1}
\eeq
with
\bea
 {\cal F}(k, n, N) \;\equiv\; \pi \lrfp{2^{3n-4} k^{2n-1}}{n {\cal G}(k,n,N) }{\frac{1}{2(n-1)}} \left(1+{\cal G}(k,n,N)^{-1}\right).
\eea
In Fig.~\ref{f} we show the function ${\cal F}(k, n, N)$ with respect
to $k$ for several values of $n$ with $N = 50$. We can see that ${\cal
  F}(k,n,N)$ is about $0.01$ for the ranges of the parameters of
interest.  This is not significantly modified for $N=40$ or $60$.

\begin{figure}[t]
\begin{center}
\includegraphics[scale=0.8]{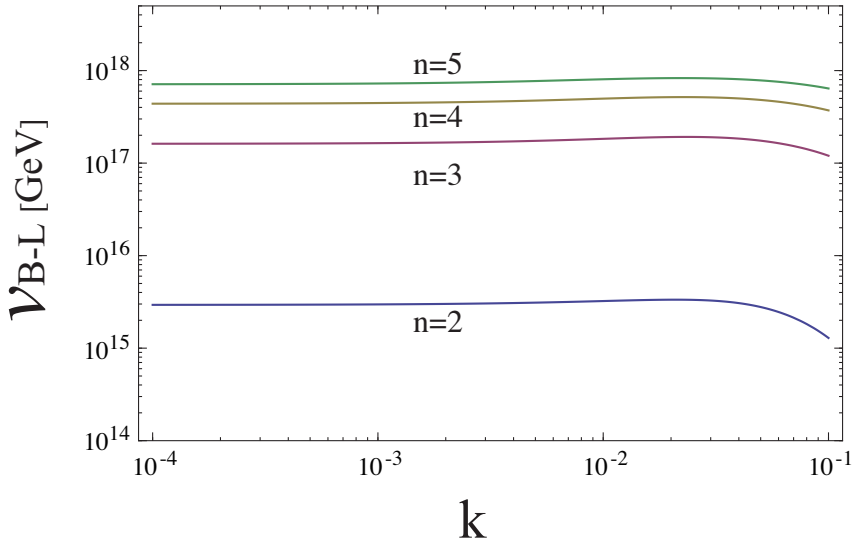}
\caption{The lower bound on the $\vbl$ as a function of $k$ for $n=2,3,4$ and $5$. We set
$N=50$.  }
\label{vbl}
\end{center}
\end{figure}

Requiring $g \lesssim O(1)$, we obtain a lower bound on $\vbl$:
\beq
\vbl \;\gtrsim\; \left(2^n \sqrt{3} \sqrt {\Delta_R^2} {\cal F}(k,n,N) \right)^{\frac{n-1}{n(2n-3)}},
\label{eq:vbl}
\eeq
which is shown in Fig.~\ref{vbl}. To be concrete we take $g = 1$ in the following analysis, and
in this case the bound on $\vbl$ is saturated. Note that the case of $n=2$ is particularly interesting because
the \ubl breaking scale is close to the see-saw scale inferred from the neutrino
oscillation.

\begin{figure}[t]
\begin{center}
\includegraphics[scale=0.8]{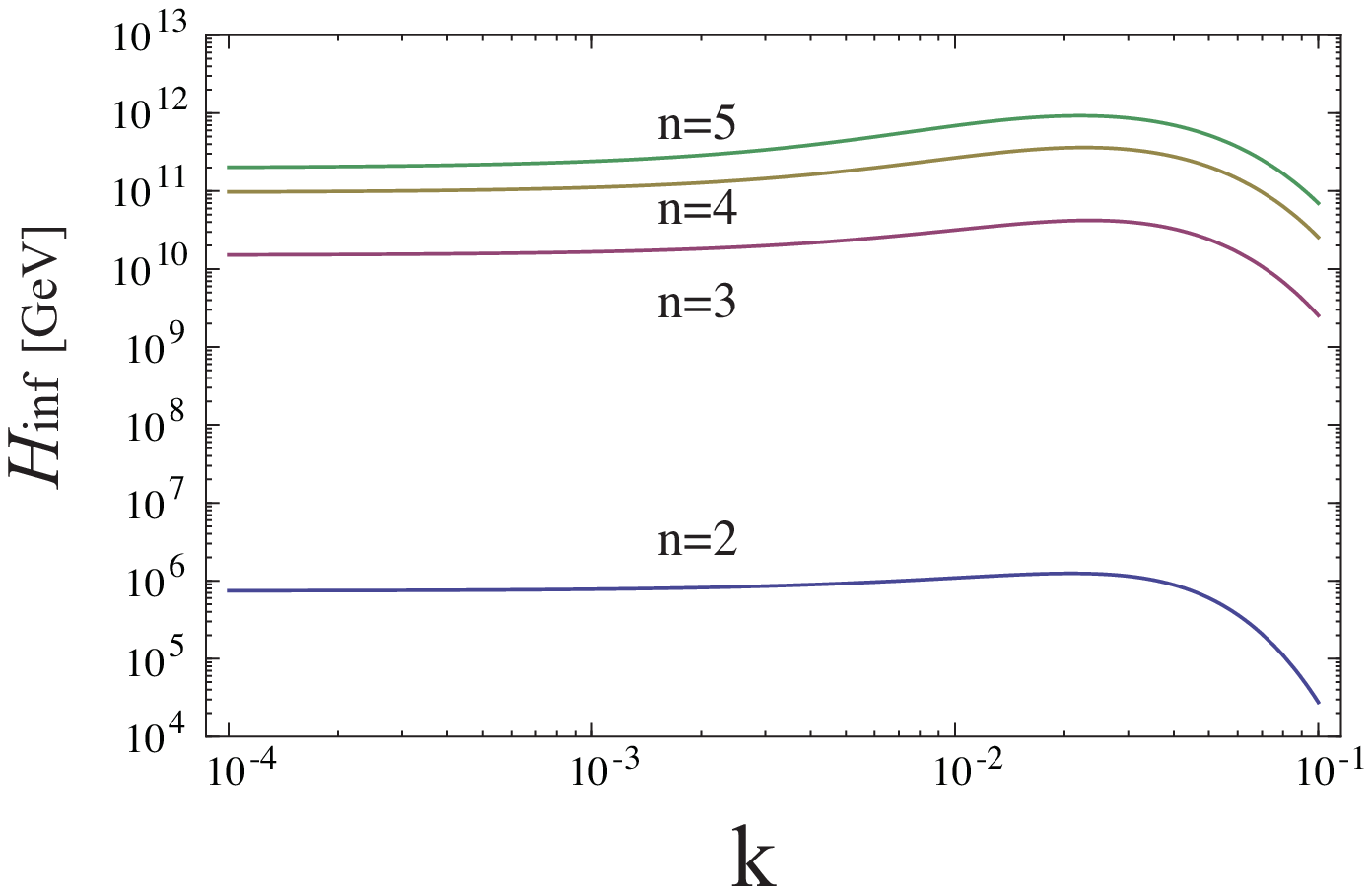}
\caption{The Hubble parameter during inflation
 as a function of $k$  for $n=2,3,4$ and $5$ and $N=50$.
$\vbl$ is given by Fig.~\ref{vbl}.}
\label{h}
\end{center}
\end{figure}

The Hubble parameter during inflation is shown in
Fig.~\ref{h}. Considering that the soft mass for the SSM particles
should be smaller than the Hubble parameter for the successful
inflation to take place, the cases of $n=2$ and $n=3$ are interesting,
especially from the point of view of
explaining the Higgs mass at around $125$\,GeV.

\begin{figure}[t]
\begin{center}
\includegraphics[scale=0.8]{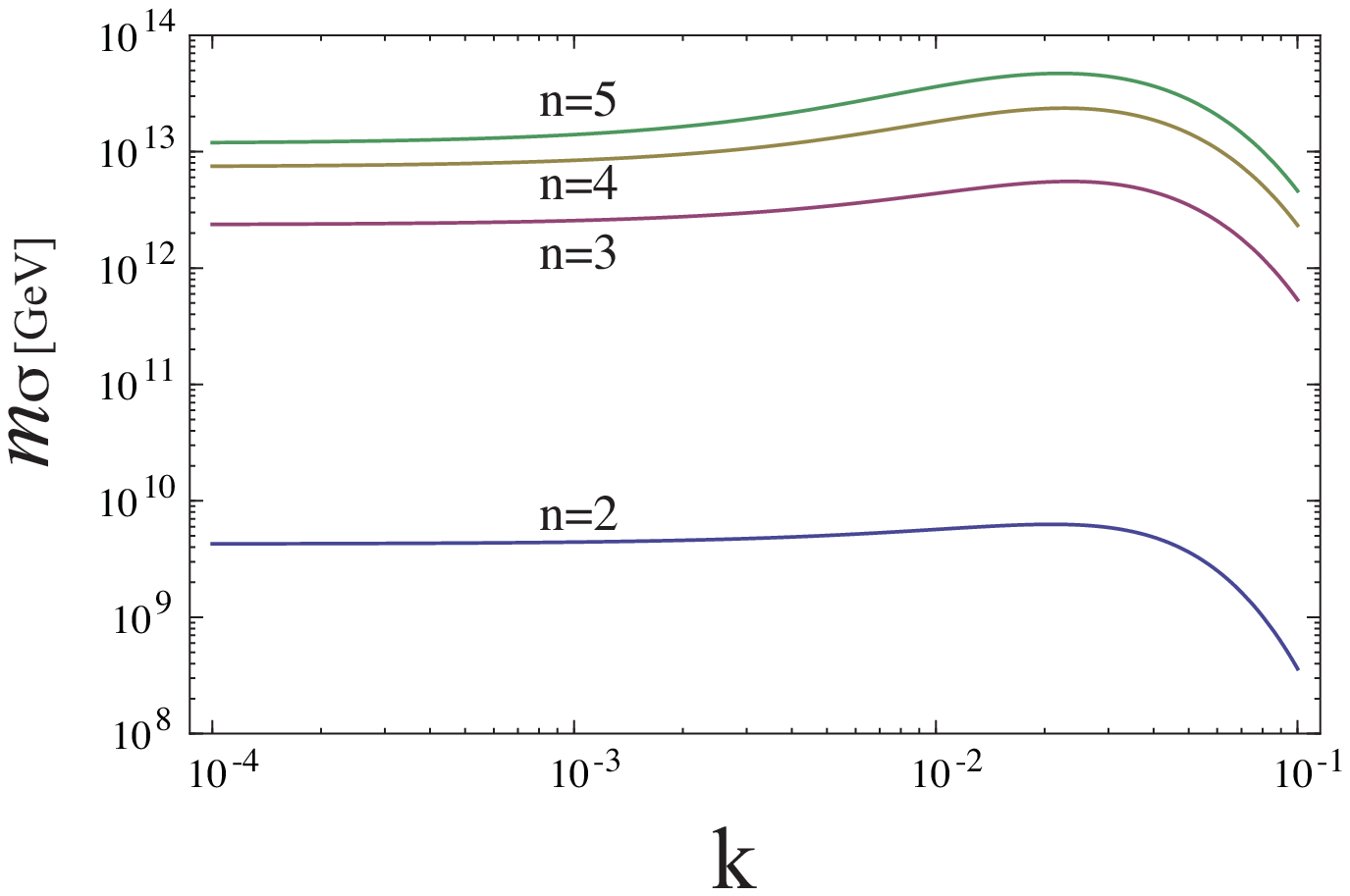}
\caption{The inflaton mass as a function of $k$  for $n=2,3,4$ and $5$ and $N=50$.
$\vbl$ is given by Fig.~\ref{vbl}. }
\label{mass}
\end{center}
\end{figure}

The inflaton mass at the potential minimum is given by
\beq
m_\sigma \;\approx\; \sqrt{2} n v^2 \lrfp{v^2}{g}{-\frac{1}{2n}} 
= \frac{2 \sqrt{3} n H_{\rm inf}}{v_{\rm B-L}} \simeq 1.7 \times 10^{-6} n 
\lrf{{\cal F}(k,n,N)}{10^{-2}} v_{\rm B-L}^{\frac{1}{n-1}}
\eeq
In Fig.~\ref{mass} we show the inflaton mass at the potential minimum
as a function of $k$ for several values of $n$. For $n\geq 3$, the
inflaton mass is greater than about $\GEV{12}$.

\begin{figure}[ht]
\begin{center}
\includegraphics[scale=0.85]{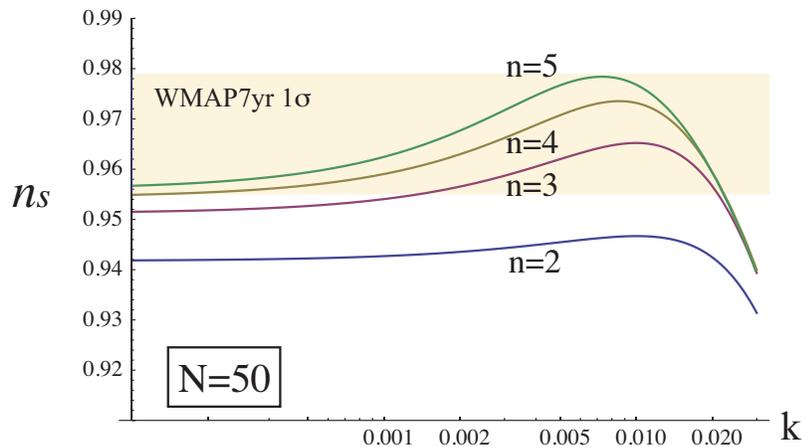}
\caption{The spectral index for $n = 2,3,4$ and $5$ in the SUSY limit.
 The shaded region shows the $1 \sigma$ allowed range,  $n_s = 0.968 \pm 0.012$,
by the WMAP 7yr data~\cite{Komatsu:2010fb}. Note that $n_s$ is independent of $\vbl$.}
\label{ns}
\end{center}
\end{figure}

Lastly let us estimate the spectral index $n_s$, which is approximately given by
\beq
n_s \;\approx 1+ 2 \eta = 1-2k \left(1+(2n-1) {\cal G}(k,n,N)^{-1}\right).
\eeq
We show the spectral index $n_s$ in Fig.~\ref{ns} as a function of
$k$.  The limit $k \rightarrow 0$ reproduces the result of
Ref.~\cite{Nakayama:2011ri}.  In principle $k$ can be extremely small,
which however requires severer fine-tuning of the parameters. If the
fine-tuning is just what is needed for the inflation to take place, we
may expect $k$ to be of $0.01$. Then, the current WMAP 7yr data $n_s =
0.968 \pm 0.012$~\cite{Komatsu:2010fb} is perfectly consistent with $n
\geq 3$, independent of the \ubl breaking scale.

\begin{figure}[h!]
\begin{center}
\includegraphics[scale=1.45]{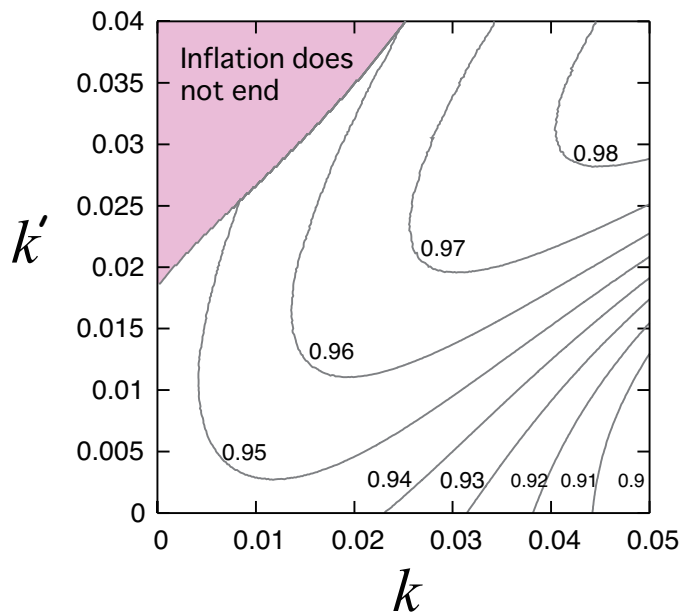}
\caption{Contours of the spectral index for $n=2$, taking account of the SUSY breaking represented by $k^\prime$
(See \EQ{vsb}). In the upper left shaded region, the inflation does not end successfully. 
Note that $k,\,k^\prime \lesssim 0.1$ must be satisfied in order for the curvature of the potential to be smaller than the Hubble parameter
for $\sigma \lesssim \sigma_{\rm end}$.
}
\label{ns-cont}
\end{center}
\end{figure}

We note that the spectral index is between $0.94$ and $0.95$ in the
case of $n=2$, which is slightly smaller than the observed value,
causing a tension at $2 \sigma$ level.  However, we should emphasize
here that the above result is derived from the potential
(\ref{Vinf}). As we discussed before, there is a finite contribution
from the CW potential once the SUSY breaking is taken into account.
Let us take account of the effect by adding the following
$V_{SB}(\sigma)$ to the inflaton potential:
\beq
V_{SB} \;=\; \frac{1}{2} k^\prime v^4 \sigma^2 \log\lrf{\s}{\s_0},
\label{vsb}
\eeq
where $k^\prime$ represents the SUSY breaking, and $\s_0$ is the
renormalization scale.  Using the result in Sec.~\ref{sec:2}, it is
given by
\beq
k^\prime \;\equiv\; \frac{1}{6\pi^2 H_{\rm inf}^2} \left(
 \sum_i  y_{\v,i}^4  m_{\tilde{N},i}^2 - 3 \gbl^2 \lrfp{q_\varphi}{2}{2} M_\lambda^2
\right).
\eeq
In order for the curvature of the potential to be smaller than the
Hubble parameter for $\sigma \lesssim \sigma_{\rm end}$, $k$ and $k^\prime$
should be smaller than $\sim 0.1$.  Note that $k$ is redefined here
so that the total potential is given by $V(\s) + V_{SB}(\s)$.  The
logarithmic correction slightly changes the global shape of the
inflaton potential, and as a result the predicted value of $n_s$ is
modified while the other inflation parameters are not significantly
changed.  We have numerically solved the inflaton dynamics and
estimated the spectral index at the pivot scale. We have fixed
$\sigma_0 = 10^{-7}$ for simplicity.  In Fig.~\ref{ns-cont}, we show
the contour of $n_s$ in the case of $n=2$, where the WMAP
normalization (\ref{bound2}) is satisfied by slightly varying the value
of $v$ and $\vbl$. The values of $\vbl$ and $H_{\rm inf}$ varies from
$3 \times 10^{15}$\,GeV to $4 \times 10^{15}$\,GeV, and $1 \times
10^{6}$\,GeV to $3 \times 10^{6}$\,GeV, respectively, in the region
shown in Fig.~\ref{ns-cont}.  We can see that $n_s$ can be increased
up to about $0.98$ in the presence of the CW correction.\footnote{
We have confirmed that $n_s$ can be increased to $\sim 0.99$ by further increasing
$k$ and $k^\prime$. This is one of the main differences from the previous works
on the two-field new inflation~\cite{Asaka:1999jb}.}  As we increase $k^\prime$, the mass near the origin
becomes more negative, while the potential becomes flatter as the inflaton goes away
from the origin. This two effects explains the behavior of $n_s$ in Fig.~\ref{ns-cont}.
We have also confirmed that the total e-folding number is greater than $100$ 
in the region shown in the figure.

Note that
$k^\prime$ = O(0.01) requires one of the right-handed neutrinos to
have a mass comparable to the inflaton VEV {\it and} that the
inequality on the SUSY breaking (\ref{bound2}) is saturated. It is
interesting that including the CW correction gives a better fit to the
observed value of $n_s$ in the case of $n=2$ where the suggested
see-saw scale of $O(10^{15})$\,GeV is close to the inflaton VEV.  Note
also that the addition of the potential (\ref{vsb}) may create a local
minimum along the inflaton trajectory, which spoil the successful
inflaton dynamics.  In order to avoid this, we demand
\begin{equation}
	k' \left[ \frac{2-n}{2(n-1)}-\log\left( \frac{\tilde\sigma}{\sigma_0} \right) \right] + k > 0,	
\end{equation}
where we have defined
\beq
\tilde\sigma \equiv 2\left( \frac{k'v^2}{2n(n-1)g} \right)^{\frac{1}{2(n-1)}}.
\eeq
This condition is violated in the upper left shaded region of
Fig.~\ref{ns-cont}.

In the case of $n=3$, there is a little hierarchy between the inflaton
VEV of $O(0.1) = O(10^{17})$\,GeV and the see-saw scale of
$O(10^{15})$\,GeV. This tension can be nicely explained by changing
the $Z_3$ assignment as $\Phi(+1)$ and $\Fb(+1)$. Then the couplings
of $\Phi$ to the right-handed neutrinos are given by
\beq
W\;=\; \frac{y_{\F, i}}{2} (\F \Fb) \F N_i N_i,
\eeq
in order to satisfy the $Z_3$ symmetry. Then we can explain this hierarchy naturally,
$10^{15}\,{\rm GeV} \sim 10^{-2} \cdot 10^{17}$\,GeV.

\section{Reheating}
\label{sec:4}
After the inflation, the inflaton must release its energy into
radiation including the SM particles, which is called the reheating.
In gauge-singlet inflation models, it is highly non-trivial if the
inflaton successfully reheats the SM sector.  In the supergravity
framework, it was shown in Ref.~\cite{Endo:2006qk} that the inflaton
is coupled to any sector via the Planck-suppressed interactions if the
inflaton has non-zero VEV, providing a robust lower bound on the
reheating temperature. At the same time, however, the inflaton would
decay into unwanted relics such as gravitinos at a non-negligible
rate~\cite{Kawasaki:2006gs,Asaka:2006bv,Endo:2007ih}, causing severe
cosmological problem.

In our present model,  the inflaton is charged under the \ubl symmetry,
and it naturally has a coupling to the right-handed neutrinos, 
\beq
W\;=\; \frac{y_{\F, i}}{2} \F N_i N_i,
\label{fnn}
\eeq
where $N_i$ denotes the right-handed neutrino chiral superfield of the
$i$-th generation, and $y_{\F, i}$ corresponds to $\sqrt{2} y_{\v, i}$
in \EQ{MN-nonSUSY}.  After inflation, $\F$ develops a VEV, and the
U(1)$_{\rm B-L}$ gets spontaneously broken. The \ubl breaking
naturally gives rise to the heavy Majorana mass $M_i \equiv y_{\F, i}
\vbl/\sqrt{2}$ for the right-handed neutrinos, as required by the
see-saw mechanism~\cite{seesaw} for the light neutrino mass.  The
above interaction induces the inflaton decay into a pair of the
right-handed neutrinos, suppressing the gravitino
production.\footnote{ In fact, low-scale inflation with a sizable
  coupling to the visible sector is favored since it suppresses the
  non-thermal gravitino
  production~\cite{Kawasaki:2006gs,Asaka:2006bv,Endo:2007ih}.  }  
The decay into the right-handed sneutrinos proceeds at the same 
rate~\cite{Endo:2006nj}. Let us comment on this process, because
it is often claimed that this decay process is suppressed compared 
to that into right-handed neutrinos. Taking the $F$-term of $\F$ in the 
interaction (\ref{fnn}) and expanding it in terms of $\chi$, we obtain
\beq
{\cal L} \;\supset - m_\sigma \frac{y_{\F, i}}{2\sqrt{2}}  \chi {\tilde N}_i {\tilde N}_i + {\rm h.c.}.
\eeq
where we have used $W_{\chi \varphi} \simeq - m_\sigma$. 
At the first sight, it seems that this interaction does not induce the inflaton
decay, however, it was shown in Ref.~\cite{Kawasaki:2006gs} that
$\chi$ and $\varphi$ gets almost maximally mixed due to the constant term
in the superpotential. The mass eigenstates are given by $(\varphi \pm \chi^\dag)/\sqrt{2}$.
Therefore, through the mixing, the inflaton decays
into the right-handed sneutrinos.

\begin{figure}[t!]
\begin{center}
\includegraphics[scale=0.7]{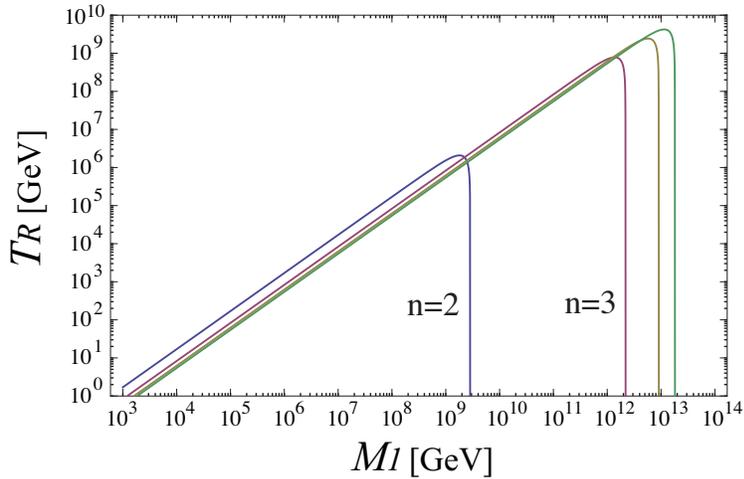}
\caption{The reheating temperature as a function of $M_1$. 
We set $g_* = 228.75$, $k=0.01$ and $N=50$.}
\label{TR}
\end{center}
\end{figure}

The decay rate of the inflaton into the lightest right-handed (s)neutrinos  is given by
\bea
\Gamma_{\rm inf}({\rm inflaton} \rightarrow N_1\,N_1,\,{\tilde N}_1\,{\tilde N}_1) & \simeq &
\frac{|y_{\F 1}|^2}{64 \pi} m_\sigma
=\frac{1}{32\pi} \frac{M_1^2 m_\sigma}{v_{\rm B-L}^2},
\eea
for $m_\s > 2 M_1$. Here we have taken account of the mixing between
$\varphi$ and $\chi$~\cite{Kawasaki:2006gs}.  The reheating temperature is
defined as
\beq
T_R \;=\; \lrfp{\pi^2 g_*}{90}{-\frac{1}{4}} \sqrt{\Gamma_\F M_p},
\label{TRdef}
\eeq
where $g_*$ counts the relativistic degrees of freedom at the
reheating.  In Fig.~\ref{TR} we show the reheating temperature as a
function of $M_1$ for $n=2,3,4,$ and $5$. We set $k=0.01$ and $N=50$,
and consider the inflation model in the SUSY limit since the effect of
the SUSY breaking on the reheating temperature is small.  Note that
the reheating temperature is so high that non-thermal leptogenesis may
work for $M_1 > \GEV{9}$ and $n \geq 2$.  We will come back
to this issue in Sec.~\ref{sec:lepto}.

\section{The SM-like Higgs boson mass}
\label{sec:5}
%
The SUSY breaking scale is bounded above by the inflationary dynamics.
If it is saturated, the typical SUSY breaking scale is of
$O(10^5)$\,GeV to $O(10^6)$\,GeV for $n=2$, and $O(10^{9})$\,GeV to
$O(10^{10})$\,GeV for $n=3$. It would be difficult to directly produce
such heavy SUSY particles at collider experiments. However, we may be
able to see a hint for such high-scale SUSY breaking from the large
radiative corrections to the SM-like Higgs boson mass.  In order to
calculate the SM-like Higgs boson mass, we need to specify $\tan
\beta$, the SUSY mass spectrum, and the stop mixing parameter, where
$\tan \b$ is the ratio of the up- and down-type Higgs boson VEVs.  In
the following we set the stop mixing parameter to be zero for
simplicity.

The possible mass spectrum can be broadly divided into the following
two cases: (1) high-scale SUSY with all the SUSY particles having a
mass comparable to ${\tilde m}$, or (2) split spectrum in
which the sfermion mass is of order $\tilde{m}$ while the gauginos and
the higgsino are at around the weak scale (or slightly higher). The
first possibility corresponds to the gravity mediation, which requires
a singlet SUSY breaking field to give a gaugino mass. The latter can
be realized in simple anomaly mediation~\cite{Giudice:1998xp} with a
generic form of the K\"ahler potential for $n=2$.  In the case of
$n=3$, we need a certain mechanism to turn off the anomaly mediation
contribution to the gaugino mass. In fact, if we take the hint for the
Higgs at around $125$\,GeV seriously, only $n=2$ is allowed for the
case (2).  Also, in the case of $n=2$, the allowed region is similar
for the cases (1) and (2).  Therefore we consider the case of (1) with
$n=2$ and $n=3$ in the following.

\begin{figure}[th]
\begin{center}
\includegraphics[scale=1.0]{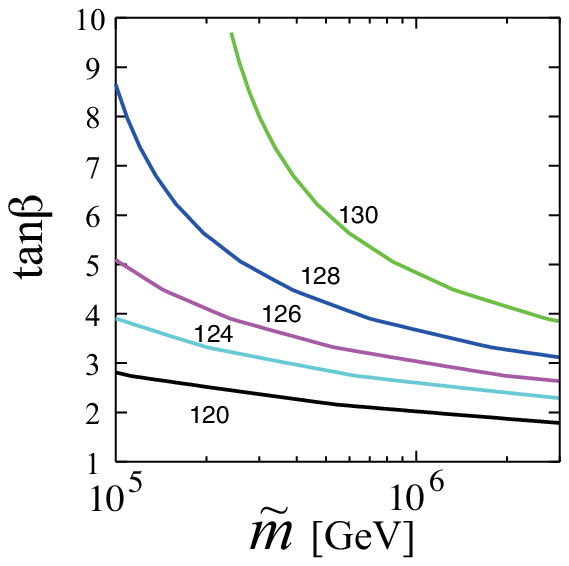}
\includegraphics[scale=1.0]{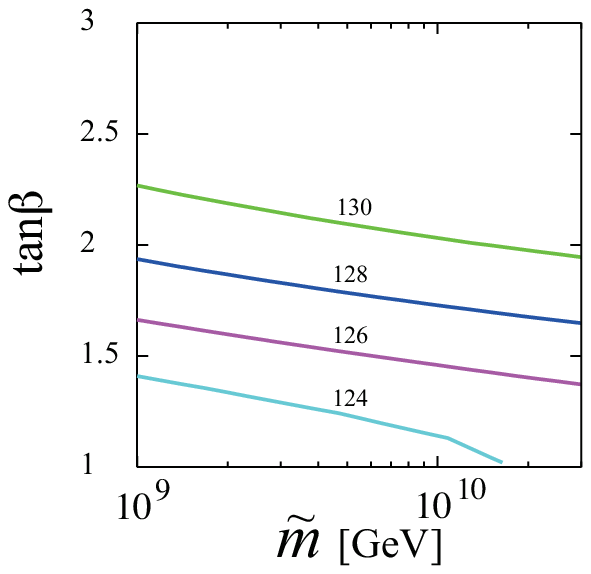}
\caption{The contours of the SM-like Higgs boson mass in the plane 
of $\tilde{m}$ and $\tan{\beta}$, corresponding to the cases of
$n=2$ and $n=3$ for which the Hubble parameter is
about $\GEV{6}$ and $\GEV{10}$, respectively.
}
\label{fig:mH}
\end{center}
\end{figure}

We have calculated the SM-like Higgs boson mass following
Ref.~\cite{Giudice:2011cg}. The contours of the Higgs boson mass $m_H$
are shown in Fig.~\ref{fig:mH}. We can see that the Higgs boson at
around $125$\,GeV suggested by the recent ATLAS and CMS experiments
can be explained for $\tan \beta = 3 - 5$ and $\tan \beta = 1 \sim1.5$
for $n=2$ and $n=3$, respectively.

The Higgs mass at about $125$\,GeV suggests a relatively high (but not 
extremely high) SUSY breaking in the minimal extension of the SSM. 
For $\tan \beta \gtrsim 1$, it varies from $10^4$\,GeV to $10^{10}$\,GeV~\cite{Giudice:2011cg}.
It is a puzzle why the SUSY should appear at such scale, which
is higher than the electroweak scale making the fine-tuning severe,
while it is much smaller than the fundamental energy scale such as the GUT or Planck scales.
Our scenario provides a possible solution to this issue: this may be due to the inflationary selection. Namely, the apparent
fine-tuning could be a result of combination of the  \ubl new inflation
and a bias toward high-scale SUSY in the landscape.

\section{Cosmological and phenomenological implications}
\label{sec:6}

In this section, we discuss various cosmological and phenomenological
implications of our scenario.  Before going further, let us briefly
mention the cosmology in the case of $n=3$.  In this case the SUSY
breaking scale is rather high: $O(10^9)$\,GeV--$O(10^{10})$\,GeV, if
the upper bound is saturated. See Eq.~(\ref{mH}) and Fig.~\ref{h}.  The
production of the SUSY particles including the gravitino is
suppressed, if the reheating temperature is (much) lower than the SUSY
breaking scale.  According to Fig.~\ref{TR}, this is the case for $M_1
\ll 10^{12}$\,GeV.  Thus there is no cosmological problem associated
with the SUSY particles.  A plausible candidate for DM will be the QCD
axion, although it may be possible that the incomplete thermalization
of the lightest supersymmetric particle (LSP) accounts for the DM
abundance.

In the following we discuss the cosmology and phenomenology, focusing
on the case of $n=2$, unless otherwise stated. Some of the discussion
below can be straightforwardly applied to the case of $n \geq 3$.

\subsection{Leptogenesis} 
\label{sec:lepto}

In the present model, the right-handed neutrinos are non-thermally
produced by the inflaton decay.  Let us see if the decay of
right-handed neutrinos can yield the right amount of the baryon
asymmetry, $n_B/s \sim 8 \times 10^{-11}$~\cite{Komatsu:2010fb}.  The
abundance of the lightest right-handed neutrino is given by
\begin{equation}
	\frac{n_{N_1}}{s} = \frac{3}{2}\frac{T_{R}}{m_\sigma}.
\end{equation}
Assuming that $N_1$ immediately decays after produced by the inflaton
decay, the lepton number generated by the $N_1$ decay
is~\cite{Asaka:1999yd}
\begin{equation}
	\frac{n_L}{s} \simeq 3\times 10^{-10}\left( \frac{T_R}{10^6\,{\rm GeV}} \right)
	\left( \frac{M_{1}}{m_\sigma} \right)
	\left( \frac{0.05\,{\rm eV}}{m_{\nu_3}} \right)\delta_{\rm eff},
\end{equation}
where $m_{\nu_3}$ denotes the mass of the heaviest left-handed
neutrino and $\delta_{\rm eff}$ the effective CP phase.  The lepton
asymmetry is related to the baryon asymmetry as $n_B/s = -(8/23)
n_L/s$ through the sphaleron process.  We find that the correct amount
of baryon asymmetry is (marginally) generated for $M_{1}/m_\sigma \sim 0.4$, $T_R
\simeq 2 \times 10^6$\,GeV and $|\delta_{\rm eff}| \simeq 1$ in the
case of $n=2$.  It is possible to enhance the baryon asymmetry in several ways. 
So far we have set $g = 1$ for simplicity. If $g \approx 0.1$, for instance,
the reheating temperature can be increased by a factor $2$ as long as $M_{1} \sim m_\sigma$.
Alternatively, if the right-handed neutrinos are degenerate, the lepton asymmetry
can be enhanced~\cite{Pilaftsis:1998pd}. The (non-)thermal
leptogenesis by the \ubl Higgs boson decay has been recently studied in detail in 
Ref.~\cite{Buchmuller:2012wn}, where the parameters are motivated by the 
hybrid inflation~\cite{Copeland:1994vg}. 

Note that the above argument assumes that there is no
additional entropy production.  Later we will show that this is indeed
the case even in the presence of the Polonyi field.  For $n \geq 3$,
the reheating temperature can be higher,  and
the right amount of the baryon asymmetry can be produced
for a broader parameter range.

\subsection{Gravitino problem}

The gravitinos are produced both thermally and non-thermally at the
reheating, and its abundance is tightly constrained by cosmology. For
$m_{3/2} \lesssim 30$\,TeV, the lifetime is shorter than about
$1$\,sec and the energetic particles produced by the gravitino decay
changes the helium-4 abundance through affecting the proton-neutron conversion
process~\cite{Kawasaki:2008qe}.  For $m_{3/2}\gtrsim 30$\,TeV, on the
other hand, the lifetime is so short that it decays before BBN, and
there is no constraint coming from BBN.  Instead, the LSPs produced by
the gravitino decay contribute to the DM density if the R-parity is
conserved.  These constraints are summarized as
\begin{equation}
\begin{split}
	Y_{3/2} \;\equiv\; \frac{n_{3/2}}{s}  \lesssim 
	\begin{cases}
	5\times 10^{-13} &~~ {\rm for}~~~10\,{\rm TeV} \lesssim m_{3/2} \lesssim 30\,{\rm TeV},\\
	\ds{4\times 10^{-13}\left( \frac{1\,{\rm TeV}}{m_{\rm LSP}} \right)}&~~ {\rm for}~~~m_{3/2} \gtrsim 30\,{\rm TeV},
	\end{cases}
	\label{gravbound}
\end{split}
\end{equation}
where $m_{\rm LSP}$ denotes the LSP mass.  Notice that the second
constraint assumes the R-parity.  If the R-parity is violated by a
small amount, the LSP can decay before BBN, and there will be no
cosmological constraint on the gravitino abundance for $m_{3/2}
\gtrsim 30\,{\rm TeV}$.

The LSP mass depends on the SUSY breaking mediation.  In the gravity
mediation we expect that the gravitino mass is comparable to the
sfermion and gauginio masses, collectively denoted by ${\tilde m}$.
If the upper bound on ${\tilde m}$ (see (\ref{mH})) is saturated, we
expect $m_{\rm LSP} = O(100)$\,TeV - $O(1)$\,PeV.  Suppose that the
LSP is the lightest neutralino.  In this case the thermal relic
abundance exceeds the observed DM density, and either late-time
entropy production or the R-parity breaking is needed. We note however
that, if the LSP mass is higher than the reheating temperature, the
LSP overproduction may be avoided.  If the gravitino is the LSP of
mass $O(100)$\,TeV, the bound (\ref{gravbound}) should read with
$m_{\rm LSP} = m_{3/2}$.  The gravitino is mainly produced by the
decay of the next-lightest supersymmetric particles, and the gravitino
abundance likely exceeds the DM density. This problem can be avoided
again by either late-time entropy production or the R-parity
violation.

In the anomaly mediation with a generic K\"ahler potential, the
gravitino mass is comparable to the sfermion mass ${\tilde m} =
O(100)\,{\rm TeV} - O(1){\rm \, PeV}$, while the gaugino mass is
suppressed (see footnote \ref{ftn}), and we expect $m_{\rm LSP} =
O(100)$\,GeV $- O(1)$\,TeV.  In the case of the Higgsino or Wino-like
LSP, its thermal relic abundance can be smaller than the observed one.

In the following we consider thermal and non-thermal production of the
gravitinos separately and show that in both cases the gravitino
abundance satisfies the cosmological bound (\ref{gravbound}).

\subsubsection{Thermal production}

Gravitinos are produced by scatterings of particles in thermal bath
during the reheating process.  The abundance is estimated to
be~\cite{Bolz:2000fu,Pradler:2006qh,Rychkov:2007uq}
\begin{equation}
	Y_{3/2}^{(\rm TP)} \simeq 2\times 10^{-16}\left(1+ \frac{m_{\tilde g}^2}{3m_{3/2}^2} \right)
	\left( \frac{T_R}{10^{6}\,{\rm GeV}} \right),
\end{equation}
where $m_{\tilde g}$ denotes the gluino mass and we have omitted the
logarithmic dependence on $T_R$ as well as terms that depend on the
other gaugino masses.  Note that the definition of $T_R$ is given by
(\ref{TRdef}).

Let us consider the case of the gravity mediation, in which the gluino
as well as the LSP have a mass comparable to $m_{3/2}$ of
$O(100)$\,TeV to $O(1)$\,PeV. The bound (\ref{gravbound}) is
marginally satisfied for $m_{\rm LSP} = 1$\,PeV and $T_R =
10^6$\,GeV. In the anomaly mediation, the bound is relaxed because of
the suppressed gaugino masses. Note that the bound disappears if the
R-parity is broken.

\subsubsection{Non-thermal production}

Gravitinos are generically produced non-thermally by the inflaton decay~\cite{Kawasaki:2006gs,Asaka:2006bv,Endo:2007ih}.
The gravitino production rate depends on the SUSY breaking mechanism. 
Let us first consider the gravity mediation. In the simple Polonyi model, there is a
singlet SUSY breaking field $z$ of mass $m_z \sim m_{3/2}$. The inflaton decays into a pair of gravitinos
through the following interaction in the K\"ahler potential
\beq
K\;=\; \frac{1}{2} (c_\F |\Phi|^2 + c_{\Fb} |\bar{\Phi}|^2) zz + {\rm h.c.},
\eeq
where $\phi$ denotes an inflaton field. The gravitino production rate is~\cite{Endo:2006tf}
\beq
\Gamma_{3/2} = \frac{1}{64\pi} {\bar c}^2 \vbl^2 m_\sigma^3,
\eeq
where we have defined ${\bar c} \equiv (c_\F + c_{\Fb})/2$.
The resultant gravitino abundance is
\beq
	Y_{3/2}^{(\rm NTP)} \sim 
	2\times 10^{-14}\, 
	\bar{c}^2
	\left( \frac{T_R}{10^{6}\,{\rm GeV}} \right)^{-1}
	\left( \frac{v_{\rm B-L}}{3\times10^{15}\,{\rm GeV}} \right)^{2}
	\left( \frac{m_\sigma}{5\times10^{9}\,{\rm GeV}} \right)^{2},
	\label{YNTP}
\eeq
where we have set $g_* = 200$.  The bound (\ref{gravbound}) can be satisfied for the LSP
mass of $100$\,TeV and ${\bar c} \lesssim 0.3$.  In the dynamical SUSY
breaking, the $z$ can have a mass much heavier than $m_{3/2}$. In this
case the gravitino production rate is similar to (\ref{YNTP}). If the
$z$ is not an elementary field but a composite one at the scale of the
inflaton mass, the gravitino production rate can be suppressed by a
factor of $O(10^2)$ or so~\cite{Endo:2007ih}. In this case the bound
(\ref{gravbound}) can be satisfied.  As mentioned before, however, the
thermal relic abundance of the LSP is generically too large in this
case. Once we introduce the R-parity violation to avoid the LSP
overproduction, there is no bound on the gravitino abundance.

In the anomaly mediation, no singlet SUSY breaking field is necessary,
and the gravitino production rate is suppressed by a factor of
$O(10^2)$ compared to (\ref{YNTP})~\cite{Endo:2007ih}. In addition, the LSP mass is
suppressed compared to the case of gravity mediation. Therefore the
bound (\ref{gravbound}) can be satisfied without introduction of the
R-parity violation, if the thermal relic abundance of the Higgsino or
Wino LSP is sufficiently small.

\subsection{Dark Matter} \label{sec:DM}
Here we discuss DM candidates in our model. Among various
possibilities, we consider the neutralino LSP and the QCD
axion. Especially in the presence of the R-parity violation, the
latter will be a plausible DM candidate, and we will study its
cosmological constraints in detail.

\subsubsection{Neutralino DM}
In the gravity mediation, the LSP has a mass of $O(100)$\,TeV or so,
and its thermal relic abundance exceeds the observed DM abundance. If
the reheating temperature is much lower than $O(100)$\,TeV, the LSP
abundance can be suppressed, which however makes it difficult for
leptogenesis to work. The simplest solution to the overabundance of
the neutralino LSPs is to break R-parity by a small amount. Then the
LSP is no longer stable, and it can decay before BBN. Of course the
LSP cannot be DM in this case, and we need another DM candidate.

In the anomaly mediation, the neutralino LSP is expected to be as
light as $O(100)$\,GeV-$O(1)$\,TeV, while the sfermion masses are much
heavier. In this case the neutralino LSP with a sizable Higgsino or
Wino fraction can account for the present DM abundance.  In fact, it
is well-known that thermal relic of the Wino LSP with mass of
$2.7$\,TeV can account for the DM~\cite{Hisano:2006nn}, while the
gravitino and scalar fermions lie at $O(1)$\,PeV.  It is intriguing
that the PeV-scale SUSY inferred from the \ubl new inflation is
compatible with the Wino DM~\cite{Nakayama:2011ri}.

\subsubsection{Axion}

Here we consider the axion cosmology.  The axion is a pseudo Nambu-Goldstone boson in association with the
spontaneous breakdown of a global U(1)$_{\rm PQ}$ symmetry, so called
the Peccei-Quinn (PQ) symmetry~\cite{Peccei:1977hh,Kim:1986ax}. The PQ
mechanism is known as the most plausible solution to the strong CP
problem in QCD. There are several ways to implement the PQ mechanism.
For simplicity we assume there is another sector in which the PQ symmetry
is spontaneously broken.
The breaking scale of the U(1)$_{\rm PQ}$ symmetry is
bounded below by the axion emission from red giant stars, and as a
result, the axion mass is extremely light. Thus the axion is stable in
a cosmological time scale, so the candidate for DM. 

In the early Universe the axion gets coherently excited, and its
 abundance is given by~\cite{Preskill:1982cy}
\begin{equation}
	\Omega_a h^2 \simeq 0.2 \left( \frac{f_a}{10^{12}\,{\rm GeV}} \right)^{1.18}\theta^2,
\end{equation}
where $f_a$ denotes the PQ symmetry breaking scale and $\theta$ the
initial misalignment angle of the axion.  Thus it accounts for the
present DM abundance for $f_a \sim 10^{11}$--$10^{12}$\,GeV without
tuning on the angle $\theta$. If we allow the fine-tuning of the
misalignment $\theta \lesssim O(10^{-3})$, $f_a$ can be increased up
to the GUT scale.

Notice that the PQ scale is higher than the inflation scale and the
reheating temperature.  Thus the PQ symmetry is likely broken already
during inflation, it is not restored after that.\footnote{ In fact
  this depends on the stabilization mechanism of the saxion.  } In
this case the axion obtains quantum fluctuations during inflation and
it contributes to the CDM isocurvature perturbation, which is
constrained by the observation of the CMB
anisotropy~\cite{Seckel:1985tj}.  Assuming that the axion is a
dominant component of DM, the magnitude of the CDM isocurvature
perturbation is estimated as
\begin{equation}
	S_{\rm c} = \frac{2\delta\theta}{\theta}= \frac{H_{\rm inf}}{\pi f_a \theta}
	 \sim 3\times 10^{-7}
	  \left( \frac{H_{\rm inf}}{10^{6}\,{\rm GeV}} \right)
	 \left( \frac{f_a \theta}{10^{12}\,{\rm GeV}} \right)^{-1}.
\end{equation}
This satisfies the observational constraint from WMAP+BAO+H$_0$~\cite{Komatsu:2010fb}:
\beq 
|S_c| \;\lesssim\; 1.4 \times 10^{-5}~~(95\%\,{\rm C.L.})
\eeq
The upper bound on $|S_c|$ will be improved by a factor $2$ or so by the Planck satellite alone. 

In the case of $n\geq 3$, the Hubble parameter is greater than
$\GEV{10}$. Therefore the axion isocurvature perturbation excludes the
axion as a DM candidate as long as the PQ symmetry is broken during
and after inflation.

Finally we comment on cosmology of the supersymmetric partners of the
axion, saxion $(s)$ and axino $({\tilde
  a})$~\cite{Rajagopal:1990yx,Kawasaki:2007mk}.  The saxion
generically obtains a mass of order the gravitino mass, and it decays
into the axion pair or the SSM particles such as gluons, Higgs boson,
and SM fermions, depending on the detailed model structure.  The
saxion is generated in a form of coherent oscillations and its
abundance is given by
\begin{equation}
	\frac{\rho_{\rm s}}{s}= \frac{1}{8}T_R\left( \frac{s_i}{M_p} \right)^2
	\simeq 2\times 10^{-8}\,{\rm GeV} \left( \frac{T_R}{10^{6}\,{\rm GeV}} \right)
	\left( \frac{f_a}{10^{12}\,{\rm GeV}} \right)^2\left( \frac{s_i}{f_a} \right)^2,
\end{equation}
where $s_i$ is the initial amplitude of the saxion. If the saxion
mainly decays into a pair of axions, its lifetime is
\begin{equation}
	\tau_{\rm s} = \left( \frac{1}{64\pi}\frac{m_{\rm s}^3}{f_a^2} \right)^{-1}\simeq
	1\times 10^{-13}\,{\rm sec}
	\left( \frac{m_{\rm s}}{100\,{\rm TeV}} \right)^{-3}
	\left( \frac{f_a}{10^{12}\,{\rm GeV}} \right)^2,
\end{equation}
where $m_{\rm s}$ denotes the saxion mass. Thus the saxion decays
before it dominates the Universe for the parameters shown in the
parentheses. Even for $s_i \sim f_a \sim \GEV{16}$, the saxion does
not dominate if $m_s \sim 1$\,PeV.

The axino  is produced thermally during the reheating~\cite{Covi:2001nw}.
Its abundance is given by\footnote{In the DFSZ axion model~\cite{Dine:1981rt,Zhitnitsky:1980tq}, 
the axino abundance is saturated for a high reheating temperature~\cite{Bae:2011jb},
and its expression is given in Ref.~\cite{Bae:2011iw} taking also account of the Higgsino decay~\cite{Chun:2011zd}.}
\begin{equation}
	Y_{\tilde a} \equiv \frac{n_{\tilde a}}{s} \simeq 2\times 10^{-7} g_s^6
	\ln \left( \frac{1.108}{g_s} \right)
	\left( \frac{f_a}{10^{12}\,{\rm GeV}} \right)^{-2}
	\left( \frac{T_R}{10^{6}\,{\rm GeV}} \right),
\end{equation}
where $g_s$ is the strong coupling constant. The axino mass depends on how the saxion is stabilized. 
Let us assume that the axino mass is comparable to ${\tilde m}$ and that the axino is unstable,
because otherwise the axino density will easily exceed the DM abundance. 
Then the axino lifetime is given by~\cite{Choi:2008zq}
\begin{equation}
	\tau_{\tilde a}=\left( \frac{\alpha_s^2}{16\pi^3}\frac{m_{\tilde a}^3}{f_a^2} \right)^{-1}\simeq
	3\times 10^{-11}\,{\rm sec}
	\left( \frac{m_{\tilde a}}{100\,{\rm TeV}} \right)^{-3}
	\left( \frac{f_a}{10^{12}\,{\rm GeV}} \right)^2,
\end{equation}
where it is assumed that the axino mainly decays into the gluino and gluon.
Hence, for $m_{\tilde a}\gtrsim O(10)$\,TeV, the axino also decays before it dominates the Universe.
There are no late-time entropy production processes from these additional particles.

\subsection{Polonyi problem}

Now we consider the cosmology of the SUSY breaking sector. In the
gravity mediation, there is a singlet SUSY breaking field $z$, so
called the Polonyi field. The Polonyi field causes a cosmological
problem as we shall briefly explain below. In the anomaly mediation,
on the other hand, such a singlet field is not necessary, therefore
there is no Polonyi problem.

First, let us review the Polonyi problem in the original
sense~\cite{Coughlan:1983ci}.  We assume that the Polonyi has only
Planck suppressed interactions, and its potential is approximated by a
quadratic potential up to the Planck scale.  The Polonyi begins to
oscillate at $H\sim m_z (\sim m_{3/2})$ and its abundance is given by
\begin{equation}
	\frac{\rho_z}{s} = \frac{1}{8}T_R\left( \frac{z_i}{M_p} \right)^2
	\simeq 1\times 10^{5}\,{\rm GeV} \left( \frac{T_R}{10^{6}\,{\rm GeV}} \right)\left( \frac{z_i}{M_p} \right)^2,
\end{equation}
where $z_i$ is the initial amplitude of the Polonyi.  Although the
Polonyi decays before BBN for $m_z \gtrsim O(10)$\,TeV, it releases a
huge amount of entropy because the Polonyi dominates the Universe soon
after the reheating.  Therefore any pre-existing baryon asymmetry is
diluted, in particular, the leptogenesis scenario does not work. In
addition, the LSPs produced by the Polonyi decay may overclose the
Universe.  One of the attractive solutions to the Polonyi problem is
to introduce an enhanced coupling between $\chi$ and the Polonyi
field~\cite{Linde:1996cx, Takahashi:2010uw, Nakayama:2011wqa}.
However, since the Hubble parameter during inflation is close to the
Polonyi mass, it is not easy to completely solve the Polonyi problem
by this mechanism~\cite{Nakayama:2011wqa}.

Second, we consider the case where the F-term of the Polonyi has a
dynamical origin~\cite{Izawa:1996pk}.  Note that the Polonyi field
itself must be an elementary singlet to give a sizable mass to
gauginos.  In this set-up, the Polonyi has a larger SUSY breaking mass
at the potential minimum, which relaxes the Polonyi problem mentioned
above. However, as noted in Ref.~\cite{Ibe:2006am}, the Polonyi field
may be driven to the Planck scale, because the potential becomes flat
at scales beyond the dynamical scale $\Lambda$.  To see this, let us
write the potential as
\begin{equation}
	V(z) \simeq \begin{cases}
		m_z^2 |z|^2 &~~{\rm for}~~|z| < \Lambda \\
		3 m_{3/2}^2 M_p^2 &~~{\rm for}~~|z| > \Lambda,
	\end{cases}
\end{equation}
where $m_z \sim \sqrt{\sqrt{3}m_{3/2} M_p}/(4\pi)^2$ is the Polonyi
mass around the origin and the dynamical scale $\Lambda$ and the
gravitino mass is related by $(\Lambda/4\pi)^2 \sim 3m_{3/2}^2 M_p^2$.
We have set coupling constants of $z$ to be order unity, for
simplicity.  In general there exists a linear term in $z$ during
inflation and it may destabilize the Polonyi field if the inflation
scale is too high.  Let us consider the K\"ahler potential $K = cz +
c^*z^*$ with constant $c$ of order $M_p$.  In the supergravity, it
yields the following term in the scalar potential during inflation
\begin{equation}
	V_{\rm lin} (z) \simeq (cz+c^*z^*)\frac{V_{\rm inf}}{M_p^2} = 3H_{\rm inf}^2 (cz +c^*z^*).
\end{equation}
In order for this linear term not to destabilize the Polonyi field, we
need $V_{\rm lin}(\Lambda) < 3m_{3/2}^2 M_p^2$.  This condition is
written as
\begin{equation}
	H_{\rm inf} \lesssim 2\times 10^8 \,{\rm GeV}
	\left( \frac{m_{3/2}}{10^{3}\,{\rm TeV}} \right)^{3/4}\left( \frac{M_p}{|c|} \right)^{1/2}.
\end{equation}
This is satisfied for $n=2$, but not for $n\geq 3$. See
Fig.~\ref{h}. The Polonyi problem is absent in the case of $n=2$ in
the the dynamical SUSY breaking scenario.  Therefore, as long as we
consider the gravity mediation, the case of $n=2$ is favored.

\subsection{Moduli stabilization and the dynamical origin of the inflation scale}

It has been known that the inflation scale should be smaller than the gravitino mass,
\beq
H_{\rm inf} \; \lesssim \; m_{3/2},
\label{KL}
\eeq
in order not to destabilize the moduli in the simple class of the
modulus stabilization models~\cite{Kachru:2003aw,Kallosh:2004yh}.  In
our scenario, the SUSY breaking is bounded above by the inflation, and
so, it is interesting to see if the inequality (\ref{KL}) can be
satisfied.

We estimated the effect of the constant term in the superpotential on
the inflaton dynamics in Ref.~\cite{Nakayama:2011ri}, assuming $|k_5|$
is of order unity, and concluded that $m_{3/2}$ should be one order of
magnitude smaller than the Hubble parameter during inflation. In fact,
this upper bound can be relaxed to be consistent with (\ref{KL}) if
$|k_5| \gg 1$ and $k_5 < 0$. Then the shift of $\chi$ due to the
constant term becomes much smaller than the Planck scale, and the
analysis so far can be applied to the case of $m_{3/2} \gtrsim H_{\rm
  inf}$. Considering that ${\tilde m}$, which is considered to be
comparable to the gravitino mass, cannot exceed $H_{\rm inf}$ (see
(\ref{mH})), the inequality (\ref{KL}) can be marginally satisfied in
our set-up, namely,
 \beq
 H_{\rm inf} \; \sim \; m_{3/2}.
\label{Hm32}
 \eeq
The reason why the successful inflation is possible even when (\ref{Hm32})
is satisfied is that the flatness of the inflaton potential is ensured by the $Z_n$
discrete symmetry in our model (\ref{KW}). This should be contrasted to
other inflation models such as the hybrid inflation~\cite{Copeland:1994vg} 
and the single-field new inflation~\cite{Izawa:1996dv} in which the inflaton is
charged under a continuous or discrete R-symmetry, and therefore the 
inflaton potential necessarily receives a correction linear in the inflaton field
once the constant term which breaks the R-symmetry is 
included~\cite{Nakayama:2010xf}.

 As mentioned before, the enhanced coupling of $\chi$ has been
 considered in context of the adiabatic solution to the moduli
 problem~\cite{Takahashi:2010uw}, and it can suppress the modulus
 abundance so that there will be no significant entropy dilution by
 the modulus decay~\cite{Nakayama:2011wqa}. This is especially the
 case if the modulus has a SUSY mass much heavier than $m_{3/2}$ as in
 the KKLT model~\cite{Kachru:2003aw}.

 Interestingly, such an enhancement naturally arises if the
 inflationary scale, $v$, in Eq.~(\ref{KW}) has a dynamical origin.
 It is straightforward to apply the IYIT model~\cite{Izawa:1996pk} to
 generate the F-term of $\chi$.  Then there is generically a coupling
 like $K \supset - |\chi|^4/\Lambda_I^2$, where $\Lambda_I$ is the
 dynamical scale. Note that since the inflaton $\Phi$ and $\Fb$ do not
 participate in the strong dynamics, there is no large contribution to
 the inflaton mass.  The dynamical scale $\Lambda_I$ is intriguingly
 close to $\Lambda$ for SUSY breaking, and so, both may be related to
 each other.

 Thus, a slight enhancement of the coupling of $\chi$, or equivalently
 lowering the cut-off scale of the $\chi$'s interaction may be the key
 to establish a successful moduli cosmology.

\section{Conclusions}
In this paper we have studied the dynamics of the recently proposed
new inflation in detail, where the inflaton is the Higgs field
responsible for the breaking of \ubl symmetry.  Importantly, we have
shown that the soft SUSY breaking is bounded above for the successful
inflation. This is because otherwise the CW potential would make the
inflaton potential too steep. Interestingly, in the case of $n=2$, the
inflaton VEV, which determines the \ubl breaking scale, is
intriguingly close to the see-saw scale of order $\GEV{15}$. The upper bound
on the SUSY breaking is then about $O(100)$\,TeV to $O(1)$\,PeV.  We have
also found that the residual CW correction can increase the predicted
spectral index  in consistent with the WMAP data.  (Note
that the spectral index will be $0.94 - 0.95$ without the CW
correction, which causes a tension at $2 \s$ level.) Furthermore we
have discussed various implications of our model: the SM-like Higgs
boson mass at about $125$\,GeV can be easily explained; non-thermal
leptogenesis works successfully; thermal and non-thermal gravitino
problem can be avoided; the DM candidates are either the lightest
neutralino (e.g. the Wino of mass $2.7$\,TeV) or the QCD axion; the
Polonyi/moduli problem can be solved; the constraint on the inflation
scale from the modulus stabilization can be marginally satisfied.
Thus, our inflation model based on the minimal B-L extension of SSM
has surprisingly many positive implications in cosmology and
phenomenology.

\section*{Acknowledgment}
This work was supported by the Grant-in-Aid for Scientific Research on
Innovative Areas (No. 21111006) [KN and FT], Scientific Research (A)
(No. 22244030 [KN and FT] and No.21244033 [FT]), and JSPS Grant-in-Aid
for Young Scientists (B) (No. 21740160) [FT].  This work was also
supported by World Premier International Center Initiative (WPI
Program), MEXT, Japan.



\end{document}